\documentclass{article}

\usepackage{PRIMEarxiv}
\usepackage[utf8]{inputenc} 
\usepackage[T1]{fontenc}    
\usepackage{hyperref}       
\usepackage{url}            
\usepackage{booktabs}       
\usepackage{amsfonts}       
\usepackage{nicefrac}       
\usepackage{microtype}      
\usepackage{lipsum}
\usepackage{fancyhdr}       
\usepackage{graphicx}       
\graphicspath{{media/}}     
\usepackage{subcaption}
\usepackage{listings}
\usepackage{xcolor}
\usepackage{amsmath}
\title{Top-P Masking for Cross Language Information Retrieval
}

\author{
  Joseph Casale\thanks{Lead author. Correspondence: \texttt{jrccasale@gmail.com}} \\
  Andrew Silverschotz \\
  Joseph DeSimone \\
}

\begin{document}
\maketitle

\begin{abstract}
  Top-K masking schemes have been proposed as a method to promote sparse representations in Information Retrieval (IR) tasks, as a simple alternative to Floating Point Operations per Second (FLOPS) regularization. Algorithms such as Bilingual Lexical and Document Expansion Model (BLADE), adopt this approach as a post-processing stage. We propose using Top-P Dynamic Masking similar to Nucleus Sampling in Large Language Models, and demonstrate better performance than Top-K masking. Specifically, we evaluate our methods in the domain of Cross Language Information Retrieval (CLIR).
\end{abstract}


\section{Important Disclaimer}
This manuscript originates from a class project conducted in 2022. The results presented are based on a partial subset of the dataset, and formal statistical significance testing has not been performed. This work has not undergone peer review, and the findings should be interpreted as preliminary and exploratory. Readers are advised to exercise caution in generalizing these results to other datasets or settings.

\section{Introduction}
 CLIR is a subdomain of classic IR where there is a language mismatch between documents and queries. Given a Query $q$ in language $\mathcal{X}$, and a set of documents $\{D\}$ in language $\mathcal{Y}$, assign a relevance ranking/score to the information expressed in the documents according to the information need expressed by the query. This problem would traditionally have two approaches: Query-Language, or Document-Language methods, where either the query is translated into the document's language, or the document is translated into the query's language. Recent approaches to this problem are referred to as Embedding-Space CLIR, where both documents and queries are transformed into multilingual embeddings and indexed, then matched accordingly with methods such as approximate nearest neighbor. Embedding-Space CLIR algorithms often use a pretrained multi-lingual transformer model, trained on any multi-lingual dataset such as Wikipedia, to generate these embeddings. Embedding-Space CLIR algorithms typically aim to promote sparsity in the representations generated, so that just the non-zero terms can be stored in the inverted indexed. 
\par
 In classic IR, SPLADE\cite{splade} generates sparse representations using a standard monolingual BERT\cite{bert}, which will be discussed in detail in Related Work. SPLADE-X\cite{spladex} and subsequently, BLADE\cite{blade}, use Top-K masking to promote sparsity in these embeddings. Top-K masking\cite{top-k} was proposed as a simple alternative to FLOPS regularization to learn sparse representations with pretrained language models. In this work, we demonstrate significant, and most important, easy-to-implement improvements in performance through the substitution of the Top-K masking scheme with our Top-P Dynamic masking scheme.

\section{Related Work}
 The SPLADE algorithm uses BERT embeddings followed by a stage called SparTerm to create a sparse vector of term importance weights. Given BERT embeddings $(h_1, h_2, ..., h_N)$ where there are $N$ tokens in the input sequence, and original input embeddings for each term in the vocabulary $V$ denoted as the set of vectors $\{E_j\}_{j=1}^{|V|}$, SparTerm results in a set of vectors $\{w_i\}_{i=1}^N$ each of size $|V|$. $w_{ij}$ may be interpreted the importance of term j in the vocabulary set for term i in the input sequence. This is done with the following neural network including a bias term $b_j$:
$$w_{ij} = LayerNorm(GeLU(LinearLayer(h_i)))^T E_j + b_j,\ \ \ j \in \{1, ..., |V|\}.$$
The term importance weights are aggregated using a RELU activation, which helps to promote sparsity with a large part of RELU's image being zero. Additionally, this also guarantees $w_j$ is a strictly positive vector:
$$w_j = \sum_{i \in terms} log (1 + ReLU(w_{i,j}))$$ 
SPLADE further promotes sparsity in $w_j$ with FLOPS regularization \cite{flopsreg} during training. Finally, document-query pairs are scored with the dot product of these vectors. 

Generalizing this method to the CLIR domain, SPLADE-X simply uses a multilingual BERT (mBERT) model, whose vocabulary encompasses more than one language. Additionally, SPLADE-X opts for the Top-K masking instead of FLOPS regularization, which works since $w_j$ is strictly positive. BLADE further improves on SPLADE-X using a pruned version of mBERT \cite{mbert} and an additional pretraining stage, while preserving other modeling choices including the use of Top-K masking. Additionally, BLADE experiments with BLADE-S, pretrained on directly translated sentences; BLADE-P, pretrained on directly translated passages; and BLADE-C, pretrained on passages of comparable meaning.  

Nucleus Sampling is a technique used in text generation proposed for Large Language Models to filter the distribution from which the next term is sampled \cite{nucleus}. We find that this technique is fundamentally very similar to Top-P Dynamic Masking, but to our knowledge, this is the first time this technique has been used for Information Retrieval. 

 \par
One significant dataset in the CLIR domain specifically is the mMARCO dataset \cite{mMarco}, which is a Multilingual version of the MSMarco \cite{MSMarco} dataset. Also, the CLEF 2003 \cite{clef} and TREC 2022/2023 NeuCLIR \cite{trecclir} tracks contain documents in many languages, such as Farsi, Russian, Chinese, Italian, German, Spanish, etc.
\par

\section{Top-P Dynamic Masking}
Consider capturing the top-p mass of the distribution of importance values, taking the maximum value from $w_j$ until attaining proportion $p$. The difference between Top-K masking and Top-P dynamic masking can be expressed mathematically as follows: 

\begin{equation*}
w_j^{(\mathrm{top}\text{-}k)} 
  = \max_{j,k} \{ w_j \}_{j=1}^{|V|}
\end{equation*}
\begin{equation}
S_p = \min \Big\{ S \subseteq V \;\Big|\; \sum_{j \in S} w_j \geq p \sum_{j \in V} w_j \Big\}, \quad
w_j^{(\mathrm{top}\text{-}p)} =
\begin{cases}
w_j & j \in S_p \\
0 & j \notin S_p
\end{cases}
\end{equation}

This formulation allows a dynamic number of terms to be selected for the input, which is either a query or a document. 

\subsection{Time Complexity}
The time complexity of Top-K masking and Top-P Dynamic masking resolve to the same tight-bound asymptote. For Top-K masking, we must sort the output vector of size $|V|$, and select $k$ elements. $k$ is chosen such that $k <= |V|$, and is thus upper-bounded by $O(|V| \log |V|)$. The time complexity of Top-K masking is $O(|V| \log |V|)$. 
Top-P Dynamic masking requires sorting the same array, but also requires summing the entire array, and selecting at most $|V|$ elements. The second and third operations are $O(|V|)$, which is also upper-bounded by $O(|V| \log |V|)$. The time complexity of this method thus also resolves to $O(|V| \log |V|)$.  

\section{Experiment}

\subsection{Research Questions}
This work is focused on answering two questions:

\flushleft{\textbf{RQ1}}: Will Top-P Dynamic Masking compared to Top-K Masking be capable of higher mAP effectiveness (as attained by high K) with higher efficiency (as attained by low K)? By allowing a dynamic number of terms to be selected by the proposed method, one may anticipate this to be the case. 

\flushleft{\textbf{RQ2}}: May there be different optimal values of P or K for the query versus the document? This seems likely because the queries are typically shorter and "simpler" than the document in terms of the possible topics that may be brought in by implicit expansion.

\subsection{Methodology}
This experiment will preserve most modeling decisions present in BLADE, just substituting the Top-K post-processing for the Top-P Dynamic masking post-processing. The BLADE authors selected $K = \lfloor 0.01 \times |V| \rfloor $ to be their default value. We use this as a reference for determining a default value to compare against. We index the NeuCLIR dataset \cite{neuclir}, specifically the 3,179,209 document mandarin subset, with the full set of NeuCLIR queries/qrels. We filter out queries without qrels before running evaluations. Due to computational limitations, our experiments are conducted by indexing the first 75000 documents in the dataset.

\flushleft{\textbf{Indexing Procedure}} The dataset is split into passages of up to 256 tokens each. We use only the queries with relevant documents. We use the topic titles for the queries, since these reflect standard query lengths. Indexing is performed using Anserini \cite{anserini}

\flushleft{\textbf{Searching Procedure}} Searching is also done with the Anserini toolkit, and scoring is done with the MaxP \cite{maxp1} operation. 

\flushleft{\textbf{Top-K versus Top-P Dynamic: }} This aspect of the experiment is focused on answering RQ1. The measurement of interest here is the trade-off between greater mAP and lower query throughput that occurs with varying levels of sparsity from both methods.
We compare the standard BLADE-C using Top-K masking to our BLADE-C using Top-P dynamic masking. This should come without loss of generality, since these methods can be easily applied and tested with any implementation. The dataset is indexed across values of $P$ between 0 and 1, and across values of $K$ between $0.005 \times |V|$ and $0.02 \times |V|$. The queries are processed using the same method, with the same corresponding $P$ or $K$ value as used for the documents. 

\flushleft{\textbf{Query versus Document: }} Here, the focus is RQ2.  Again, the measurement of interest is the trade-off between mAP and query throughput, but we analyze the results of choosing different values of $P$ for queries than we do for documents. Specifically, we try all the possible combinations of P=0.25, 0.35, 0.45, 0.55, 0.65, 0.75, 0.85, 0.95, 0.99, excluding the combinations assigning the same value of $P$ for both documents and queries, for a total of 72 trials. This experiment is also run with BLADE-C.

\subsection{Implementation}
 Experiments are implemented with the PyTorch\cite{pytorch} autograd package. Indexing and searching is done using Anserini\cite{anserini}. The metrics are computed with the Python package ir measures \cite{irmeasures}. All experiments are performed with one NVIDIA RTX 1080 GPU. 

\section{Results}
In this section, we analyze the performance of Top-K and Top-P-Dynamic masking by comparison of obtained mAP scores and their associated query throughput. Total indexing time is computed as the sum of Anserini's indexing, and PyTorch-hosted inference code. All experiments were performed using an Nvidia GTX 1080 GPU.

\subsection{Analysis}
\begin{figure}[h!]
    \centering
    \begin{subfigure}{0.45\textwidth}
        \includegraphics[width=\linewidth]{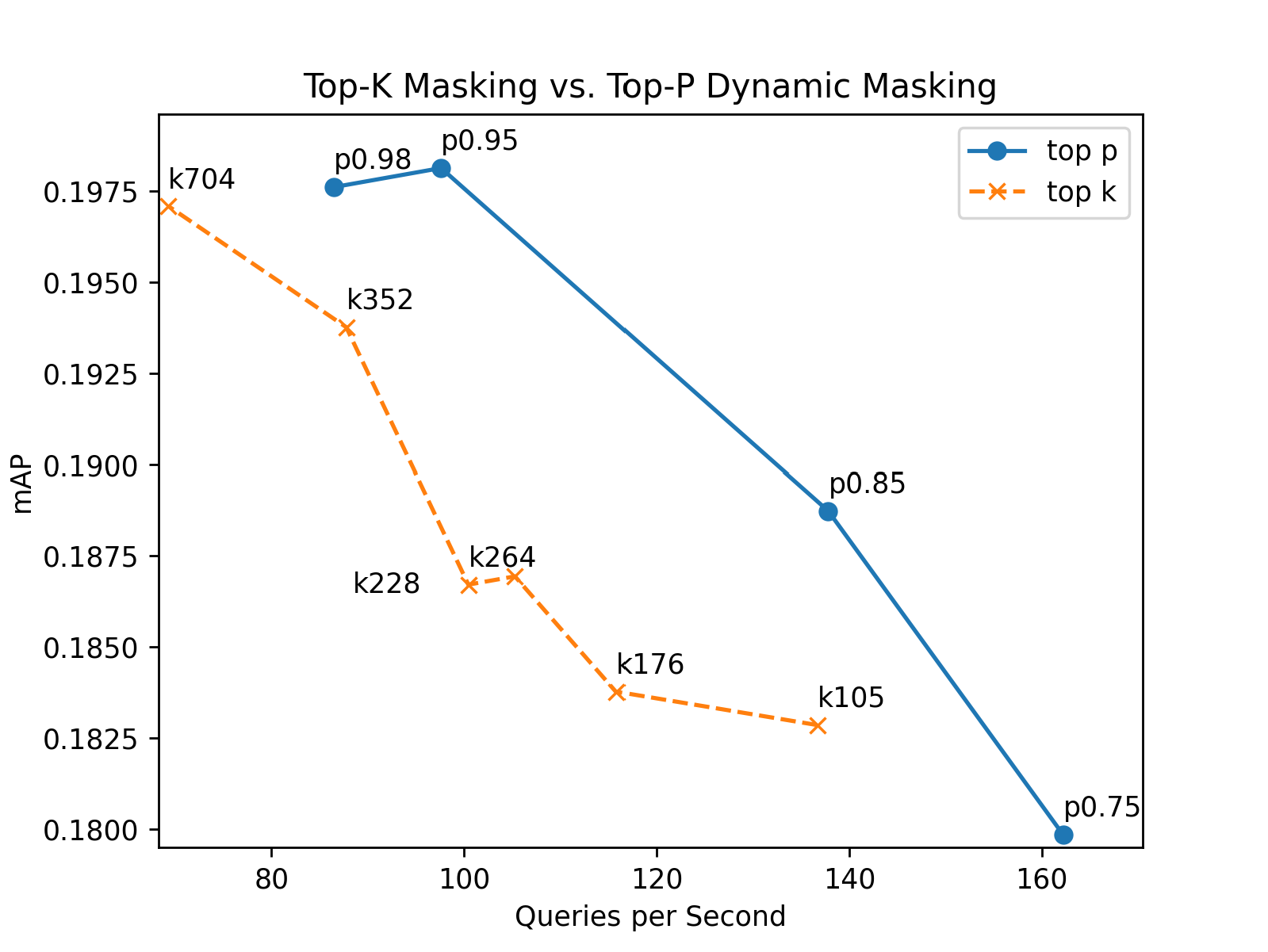}
        \caption{mAP vs query throughput for different values of $k$ and $p$}
    \end{subfigure}%
    \hspace{0.05\textwidth} 
    \begin{subfigure}{0.45\textwidth}
        \includegraphics[width=\linewidth]{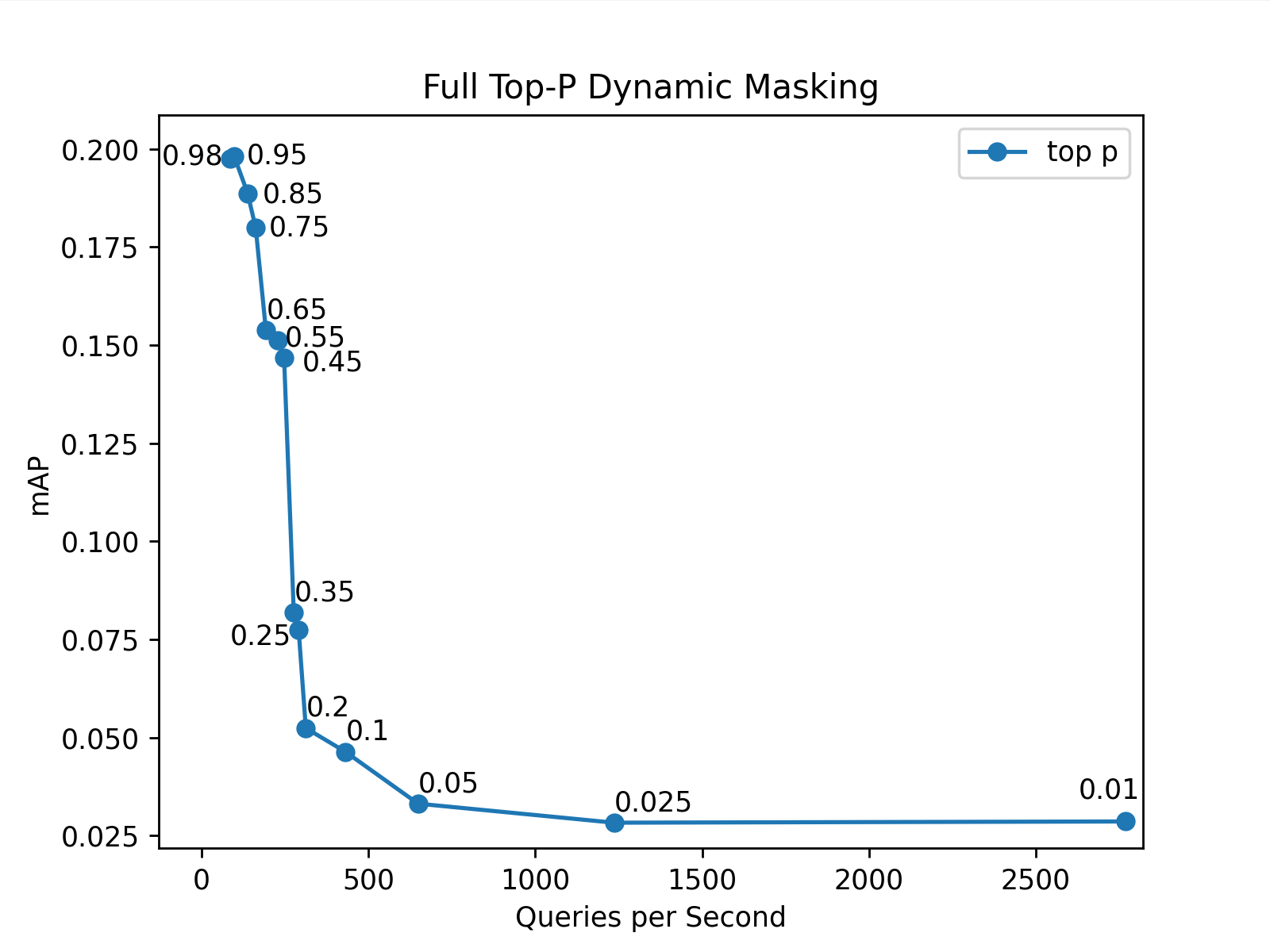}
        \captionsetup{width=0.85\linewidth}

        \caption{mAP vs query throughput for full range of $p$ tested.}
    \end{subfigure}
    \caption{Results of experiment comparing Top-K masking to Top-P-Dynamic masking, re-indexing the small dataset.}
    \label{fig:pvskres}
\end{figure}

First, we reproduce the results of BLADE-C's efficiency vs effectiveness analysis by varying values of $k$. Next, implementing Top-P-Dynamic masking in place of Top-K, and varying $p$ in the range \((0, 1)\), we produce similar results. The final results from running with Top-P-Dynamic masking and Top-K masking on 75,000 documents suggests that Top-P-Dynamic masking consistently outperforms the Top-K masking method. Figure \ref{fig:pvskres}.(a) shows that, while these score differences are relatively minor, they seem to be consistent across comparable values of $p$ and $k$. 
For Top-P-Dynamic masking, the more restrictive values will process more queries more quickly at the cost of precision. We ran a variety of $p$ values to demonstrate how mAP and queries per second change across the range of values chosen. Figure \ref{fig:pvskres}.(b) shows that mAP quickly drops when reducing $p$ from a large value. 

\begin{figure}[h!]
    \centering
    \begin{subfigure}{0.45\textwidth}
        \includegraphics[width=\linewidth]{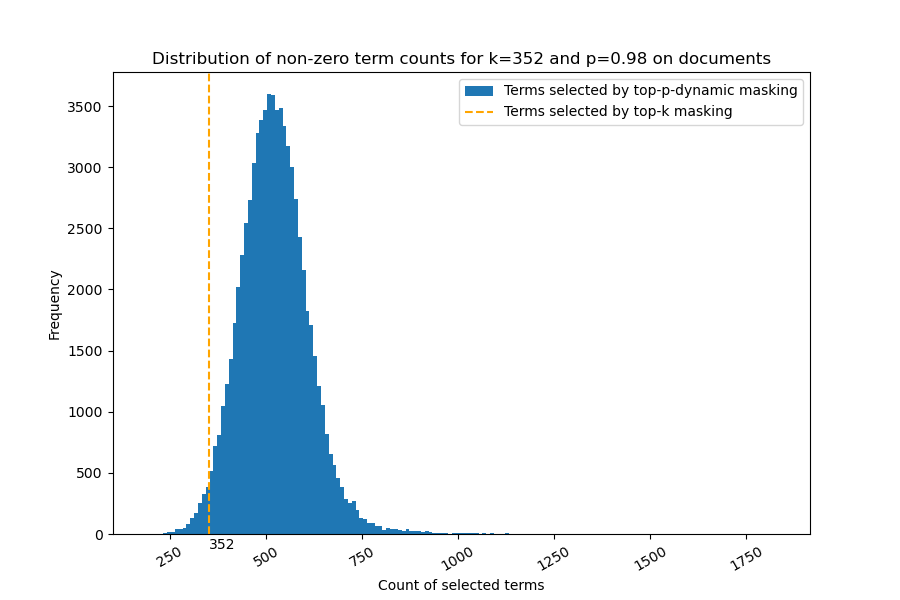}
        \caption{Distribution of terms selected for each document. On average, more terms were selected for a document.}
    \end{subfigure}
    \hspace{0.05\textwidth} 
    \begin{subfigure}{0.45\textwidth}
        \includegraphics[width=\linewidth]{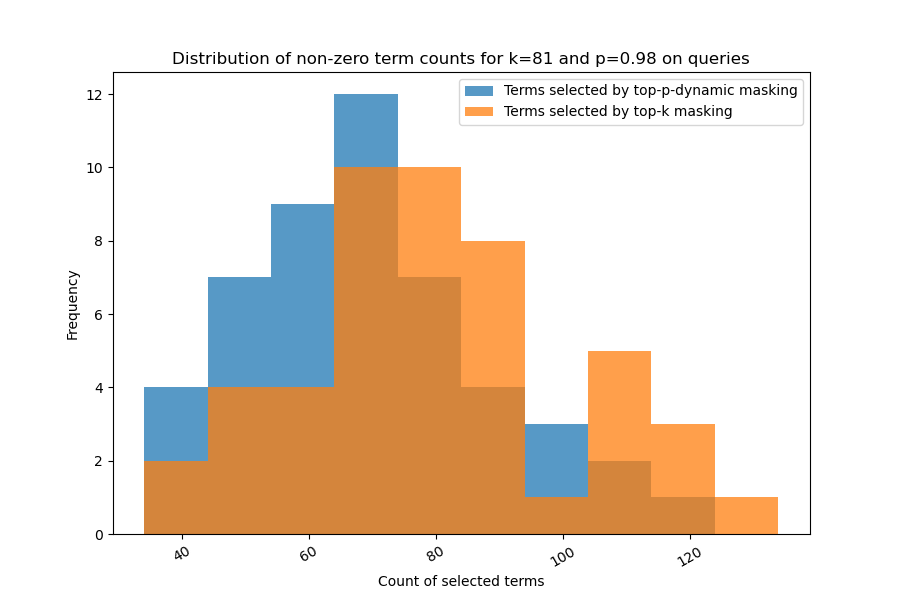}
        \caption{Distribution of terms selected for each query. On average, fewer terms were selected for a query.}
    \end{subfigure}%
    \captionsetup{width=1\linewidth}
    \caption{Distributions of terms selected by Top-K masking and Top-P-Dynamic masking for queries and documents, fixing $k=352$ and $p=0.98$. For Top-K masking on documents, the upper limit of 352 terms was chosen for every document.}
    \label{fig:pvsk_totals}
\end{figure}

The $p$-value of 0.85 seems to have a particularly large improvement in query throughput compared to Top-K masking at similar mAP scores, but this trend appears to be overall present. To demonstrate why this may be, we plot the distribution of the number of terms selected by each method. Since the original authors of BLADE set their default value of $k$ to 352, we choose to set $p$ to 0.98, since they were very close in query throughput. 
\\ \quad
Figure \ref{fig:pvsk_totals} shown below investigates the difference in mAP with similar results in query throughput for comparable $p$ and $k$ values. Figure \ref{fig:pvsk_totals}.(a) displays that a reasonably-sized document will contain enough non-zero terms to reach the ceiling $k$ easily, while top-p dynamic masking will generally select more terms. However, as can be seen in \ref{fig:pvsk_totals}.(b), when we plot the distributions of terms selected for each query, the ceiling of $k$ is not reached, allowing for all non-zero terms to be selected. Meanwhile, the terms selected by Top-P-Dynamic masking will still be only a proportion of total weight, resulting in Top-K masking selecting more terms on average for queries. This selecting of more terms for documents but fewer for queries appears to result in a greater mAP for a similar throughput.

\begin{figure}[h!]
    \centering
    \includegraphics[width=0.75\linewidth, height=0.35\linewidth]{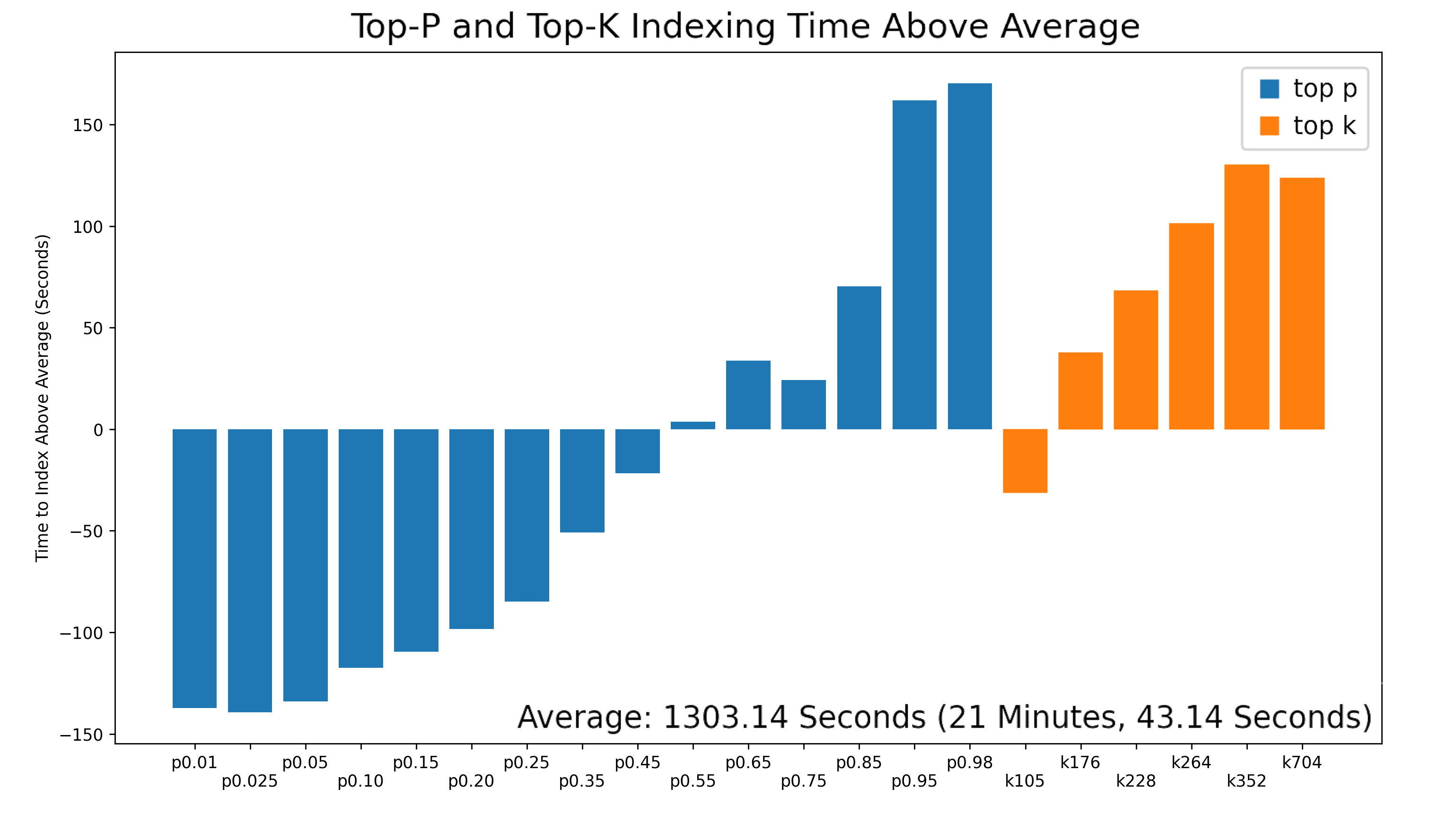}
    \captionsetup{width=0.65\linewidth}
    \caption{The Time Above Average for each Top-P and Top-K Method.}\label{fig:indextime}
\end{figure}
As expected, the indexing time increases when higher amounts of words are selected for each document. This once again supports the increase in query latency when the $p$ value is increased to include more terms. In  addition, the comparable $p$ value of 0.98 is indexed for about 20 seconds longer than its counterpart $k$, which is also explained by the general difference in the terms of the selected document in \ref{fig:pvsk_totals}.(a). We also provide the total indexing time for the entire NeuCLIR dataset of size 3,179,209 documents. 
\vspace{10pt}

\centering{
\begin{tabular}{|c|c|c|}
     \hline
      & Top-K & Top-P-Dynamic \\
     \hline
     Total Index Time & 10 hr, 19 min& 11 hr, 02 min\\
     \hline
\end{tabular}}

\newpage

\bibliographystyle{ACM-Reference-Format}
\bibliography{references}

\end{document}